\documentclass[[preprint,twocolumn,preprintnumbers,amsmath,amssymb,superscriptaddress,noshowpacs]{revtex4-1}

\usepackage{graphicx}
\usepackage{dcolumn}
\usepackage{bm}

\graphicspath{{COLOR/}}

\begin{document}

\title[Propagation dynamics of successive emissions in laboratory and astrophysical jets and problem of their collimation
]{Propagation dynamics of successive emissions in laboratory and astrophysical jets and problem of their collimation
}

\author{I. Kalashnikov}
	\email{kalasxel@gmail.com}
	\affiliation{National Research Nuclear University MEPhI, 31 Kashirskoe shosse, Moscow 115409, Russia.}
    \affiliation{Institute for Computer Aided Design, Russian Academy of Sciences, 19/18 2-ya Brestskaya st., Moscow 123056, Russia.}	
	
\author{P. Chardonnet}
	\affiliation{National Research Nuclear University MEPhI, 31 Kashirskoe shosse, Moscow 115409, Russia.}
	\affiliation{Univ. Grenoble Alpes, USMB, CNRS, LAPTh, 9 chemin de Bellevue, F-74941 Annecy-le-Vieux, France.}
	
\author{V. Chechetkin}
	\affiliation{National Research Nuclear University MEPhI, 31 Kashirskoe shosse, Moscow 115409, Russia.}
	\affiliation{Institute for Computer Aided Design, Russian Academy of Sciences, 19/18 2-ya Brestskaya st., Moscow 123056, Russia.}
	\affiliation{National Research Center ''Kurchatov Institute'', 1 Akademika Kurchatova pl., Moscow 123182, Russia.}
	\affiliation{Keldysh Institute of Applied Mathematics, Russian Academy of Sciences, Moscow 125047, Russia.}

\author{A. Dodin}
	\affiliation{Sternberg Astronomical Institute, Moscow M.V. Lomonosov State University, Universitetskij pr., 13,  Moscow, 119992, Russia.}

\author{V. Krauz}
	\email{Krauz\_VI@nrcki.ru}
	\affiliation{National Research Center ''Kurchatov Institute'', 1 Akademika Kurchatova pl., Moscow 123182, Russia.}


\begin{abstract}
The paper presents the results of numerical simulation of the propagation of a sequence of plasma knots in laboratory conditions and the astrophysical environment. The physical and geometric parameters of the simulation have been chosen close to the parameters of the PF-3 facility (Kurchatov Institute) and the jet of the star RW Aur. We found that the low-density region formed after the first knot propagation plays an important role for collimation of the subsequent ones. Assuming only the thermal expansion of the subsequent emissions, qualitative estimates of the time taken to fill this area with the surrounding matter and the angle of jet scattering have been made. These estimates are consistent with observations and results of our modeling.
\end{abstract}

\pacs{97.21.+a; 52.58.Lq.}
\keywords{Young stellar jets, Laboratory plasma astrophysics, Supersonic flow, Shock waves, Jet-ambient interaction, Collimated flows, Radiative cooling, Numerical simulations.}

\maketitle

\section{introduction}

Cumulative observational evidences have shown that high collimated jets is an universal phenomena not only  in quasars and microquasars but also in young stellar objects (YSOs).

The Herbig-Haro objects were the first direct evidence of a bright bow shock as the result  the collision of such jets with the external gas surrounded the star \cite{Herbig51,Schwartz}. In the 1980s, the observations of many bipolar outflows from young stars  definitely confirm  the previous observations\cite{Herbig81,Bally06}. Since then, high angular resolution observations  and wide wavelenght coverage show the motion of jet and the measurement of the kinematics parameters \cite{Hirth97}. In particular subsecond angular resolution images from  Hubble Space Telescope allow to clarify the nature of bight knots in the jets \cite{Reipurth89}.

Now, it is clear that jets and outflows from YSOs  play  a key role in the process of star formation.  
The current picture is that they evade angular momentum during formation of stars and protoplanetary disks.  Although, we have not the complete theory, it appears that the process of jet collimation and the accretion disks are the ingredients of all modelisation of YSOs.  There is a consensus that  the jet formation is related to a poloidal magnetic field anchored in the disk \cite{Ferreira97}.

Turbulent accretion in the disk may lead to the angular momentum transport via magnetorotational instabilities.  Therefore numerical simulations are the other essential ingredient  toward the understanding of the all process.

 In 2005, the production of  supersonic plasma jets using Z-pinch facilities has been shown by S. V. Lebedev et al. \cite{Lebedev}. The main idea is to use the scaling law in magnetohydrodynamics  to simulate jet propagation in laboratory with Z-pinches and high powerful lasers  \cite{RyutovandRemington,Remington,Ryutov}. This field is certainly becoming a new era of astrophysical laboratory with promissing results \cite{Abertazzi,Belan}.
 
 Here we will present a new complementary approach in the quest of understanding jet collimation based on a numerical simulations of knots propagation confront to experiment in laboratory \cite{Krauz1,KrauzUFN,Krauz2,Krauz3,Krauz4,Krauz5}. This facility is of the ''plasma focus'' type (PF), which is one of the modifications of Z-pinches - the so-called "non-cylindrical Z-pinch".  One of the advantages of the experiment scheme on the PF-units in comparison with the most well-known analogs (lasers, fast Z-pinches) is the possibility of modeling the propagation of a plasma jet over sufficiently large distances, which allows us to investigate the dynamics of jet parameters, such as density, temperature, the distribution of magnetic fields in its interaction with the surrounding medium. The scheme of the experiment is described in more detail in the works \cite{Krauz1,Krauz3}. Our work presents interesting possibilities for self-collimation of jets.

\section{Model setup}

\subsection{Scaling law and dimensionless numbers}

The idea of scaling law is to create a laboratory plasma that is scale model of an astrophysical one. The conditions for hydrodynamic similarity have been analyzed by Ryutov \cite{Ryutov}. This implies three assumptions: a collisional medium, no viscosity and the energy flow by radiation or conduction should be negligible.

The collisional medium implies that the mean free path $\bar{l}$ and the Larmor radius $r_L$  satisfy the conditions   $\bar{l} \ll L$ and $r_L \ll L$  where $L$ is the typical size of our system.

The second condition requires that the Reynolds number which represents the ratio of the inertial force to the viscosity satisfies: $Re \gg 1$.  We  will estimate the Reynolds number in a partially ionized plasma. The dynamic viscosity is a result of interaction of ions and atoms: $\eta \simeq \rho \bar{v}\bar{l}$, where $\rho$ is density. Thus, the Reynolds number may be written as $ Re = V L / \bar{v}\bar{l} $, where $V$ and $L$ are characteristic velocity and size of a flow. The mean free path of a particle path is expressed as $ \bar{l} = n \bar{v} / N $, where the total number of collisions of these particles  per  volume and time is $N = n^2_i \overline{\sigma_{ii} v_{ii}} + n^2_n \overline{\sigma_{nn} v_{nn}}+n_i n_n \overline{\sigma_{ni} v_{ni}}$, with $n_i$, $n_n$ - the concentrations of ions and atoms respectively, $\sigma_{ii}$, $\sigma_{nn}$, $\sigma_{ni}$ - the corresponding scattering cross sections and $v_{ii}$ $v_{nn}$ $v_{ni}$ - modules of relative velocities, the bar denotes averaging over a speed.  We will assume that all particles move with almost the same velocity. Also let's introduce the degree of ionization: $\alpha = n_i/(n_i+n_n)$, believing that all atoms are singly ionized. Then the average length of a particle path may be expressed as $ \bar{l} = 1/(n(\alpha^2\sigma_{ii} +(1-\alpha)^2\sigma_{nn}+\alpha(1-\alpha)\sigma_{ni} )) $. Further, suppose that all neutral atoms scatter on other neutral atoms and ions equally i.e. $\sigma_{nn}\simeq\sigma_{ni}$. The average thermal velocity can be expressed in terms of temperature: $\bar{v}\simeq \sqrt{kT/m_i}$, $m_i$ is mass of a ion or, equivalently, mass of an atom, $k$ is the Boltzmann constant. The ion-ion scattering cross section is well known - this is Rutherford's formula: $\sigma_{ii} = {4\pi(Ze)^4}\Lambda/{(kT)^2}$, where $Z$ is an atomic number, $e$ is the charge of an electron and $\Lambda$ is the Coulomb logarithm. The cross section of atom-atom scattering is supposed to be area of a circle with atom radius: $\sigma_{ni}\simeq\sigma_{nn}=\pi r^2_a$. Finally, we obtain the following expression for the Reynolds number:
\begin{equation}
\label{Re}
 Re = \sqrt{\frac {m_i} {kT}} \left( \alpha^2 \frac{4\pi(Ze)^4}{(kT)^2} \Lambda + (1-\alpha)\pi r_a^2 \right) n V L.
\end{equation}

 The magnetic Reynolds number is given by $ Rm = 4\pi\sigma V L / c^2 $ where $\sigma$ is the conductivity of the plasma, $c$ is the speed of light.  Because of Ramsauer effect the scattering of electrons with energies of the order of $ 1 \text{eV} $ on neutral atoms does not almost occur, therefore the conductivity of the plasma is affected only by the ions. In this case the mean free path of electrons equals $\bar{l} = 1/\alpha n \sigma_{ei}$, where $\sigma_{ei}$ is the cross section of electron-ion scattering, which equals $\sigma_{ie} = {4\pi(Ze^2)^2}\Lambda/{(kT)^2}$. The conductivity of plasma may be written as $\sigma \sim e^2 \alpha n \bar{l} / m_e \bar{v}$. Then we get the value of the Reynolds magnetic number, not depending on the degree of ionization:
\begin{equation}
\label{Rm}
Rm = r_0 \sqrt{m_e} \frac{(kT)^{ 3 /2}}{(Ze^2)^2 \Lambda } V L,
\end{equation}
where $r_0$ is the classical electron radius. 

The third condition is related to the Peclet number which represents the ratio of the heat convection to the heat conduction: $P_e = L V/ \kappa$.
In the plasma, the heat transfer is done essentially by the electrons and the coefficient of heat transfer is estimated as $\kappa \sim C_e \alpha n \bar{l} \bar{v} $, where $C_e\sim 1$ is electron heat capacity. Therefore we obtain the expression: 
\begin{equation}
\label{Pe}
 Pe = \frac{4\pi(Ze^2)^2}{(kT)^{5/2}} \Lambda \sqrt{m_e} \alpha n V L.
\end{equation}

Another interesting number is the Mach number simply given by:

\begin{equation}
\label{M}
 M=  \frac{V}{c_s}  = V\sqrt{ \frac{m_i}{\gamma kT} },
\end{equation}
where $\gamma$ is heat capacity ratio.  

\begin{table*}
\caption{\label{tabl:Param} The parameters of the astrophysical and laboratory jets.}
\begin{center}
\begin{tabular}{lcccc}
\hline\hline
 \ Parameter 		& {\ RW Aur (red)} 	& \ RW Aur (blue) & \ PF-3 (H) 	& \ PF-3 (Ar) \\
\hline
\ Reynolds number, $Re$ & \ $10^6$ 	&\ $10^8$ 		&\ $10^2-10^4$ &\ $ 10^3-10^5$ \\
\ Magnetic Reynolds number, $Rm$ &\ $10^{14}$ &\ $10^{15}$ &\ $18$ & \ $18$ \\
\ Peclet number, $Pe$ &\ $10^6$ &\ $10^7$ &\ $10-10^2$ &\ $10-10^2$  \\
\ Mach number, $M$ &\ $ > 1$ &\ $> 1$ &\ $4$ &\ $25$ \\
\ Internal Mach number, $M_*$ &\ $20$  &\  $18$ &\ $1.7$ &\ $11$ \\
\ Euler number, $Eu$   & $14.9$ &\ $23.2$ &\ $2.3$ &\ $14.5$ \\
\ Ratio of densities, $n_{\text{jet}}/n_{\text{ambient}}$ &\ $>1$ &\  $>1$ &\ $3-5$ &\ $3-5$ \\
\hline\hline
\end{tabular}
\end{center}
\end{table*}

Provided the three conditions are satisfied, it can be shown that the two hydrodynamical systems in lab and in astrophysics are similar and we can use the following relation\cite{Ryutov}:
\begin{equation}
\label{Euler}
Eu= v_\text{lab} \sqrt{\frac{\rho_\text{lab}}{p_\text{lab}}}  = v_\text{astro} \sqrt{\frac{\rho_\text{astro}}{p_\text{astro}} },
\end{equation}
where $Eu$ is the Euler number.  Therefore, if we know the timescale of jet lab experiment, we can deduce the time scale for jet in astrophysical environment using the relation:
\begin{equation}
\label{timescale}
\tau_\text{astro} = \tau_\text{lab} \frac{L_\text{astro}}{L_\text{lab}}  \sqrt{ \frac{ p_\text{lab} / \rho_\text{lab}}{ p_\text{astro} / \rho_\text{astro}  } }   
\end{equation}

\subsection{Initial conditions}

We have used a 2D cylindrical setup with $r$ and $z$ as variables. The magnetic field have been set only by the azimuthal: it grows linearly from $r=0$ up to jet bound and decrease as $r^{-1}$ from the bound. For the laboratory jet the value of the magnetic field at the bound is $B_0 \sim 1\text{ kG}$, according to the experiments \cite{KrauzUFN}. The magnetic field at the ending of the plasma cylinder is believed to break off. 
Temperatures of jet and its ambient at initial moment have been set by not very different. 
In case of this choosing of thermodynamical values all observed parameters are achieved after several steps of calculation. Such configuration of the magnetic field corresponds to the measurements, carried out on the PF-3 facility \cite{KrauzUFN}. Although this formulation of the problem does not take into account reverse closing currents and real direction of vector of magnetic induction, it allow us investigate joint influence of the azimuthal magnetic field, radiation cooling and internal medium and also a role of a canal, formed by a first ejection, on the collimation of subsequent ejection.

The laboratory jet propagates with the following parameters\cite{KrauzUFN}: $n\simeq10^{17} \text{ cm}^{-3}$, $T\simeq 5 \text{ eV}$, $V\sim 5\cdot 10^6 \text{ cm/s}$. The degree of ionization usually is unknown and changes from point to point, therefore it stays as free parameter. The plasma in the facility is supposed to be weakly ionized, therefore in each cases we take $Z=1$. The density of the ambient is less than jet density in $3-5$ times and its temperature is $T\simeq 1\text{ eV}$.

Regarding the astrophysical parameters for the YSOs, we have chosen the the red jet of RW Aur star\cite{Astro}. We consider a plasma of hydrogen with the  following parameters: $n=65900 \text{ cm}^{-3}$, $T=1.06\text{ eV}$, $\alpha=0.08$, $V=1.5\cdot 10^7 \text{ cm/s}$. The blue has: $n=22900 \text{ cm}^{-3}$, $T=1.42\text{ eV}$, $\alpha=0.23$. $V=2.7\cdot 10^7 \text{ cm/s}$. Regarding the parameters surrounding these jets environment we cannot say something specific, therefore it could only be stated that $M>1$ and $n_{\text{jet}}/n_{\text{ambient}}>1$.

The table \ref{tabl:Param}  summarizes the using values of the main parameters, calculated with formulas (\ref{Re}) - (\ref{M}). It can be seen, that  hydrodynamically the red jet of RW Aur star and laboratory experiments with argon are most similar. The difference between their magnetic Reynolds number in the context of our model is apparently not important, because we do not investigate the effect of energy dissipation because of the flow of electric currents. It is enough for us that all these numbers are much greater than unity. 


\subsection{Simulation setup}

Following the arguments on the scaling law developed in II A, the equations of the evolution of such plasma are described in the frame of ideal MHD:
\begin{equation}
 \frac{\partial\rho}{\partial t} + \operatorname{div} \rho \mathbf{v} = 0,
\end{equation}
\begin{equation}
 \rho\frac{\partial\mathbf{v}}{\partial t} + \rho(\mathbf{v},\nabla)\mathbf{v} = - \nabla p -\frac{1}{4\pi} [\mathbf{H} \times \operatorname{rot} \mathbf{H}],
\end{equation}
\begin{equation}
 \frac{\partial\mathbf{H}}{\partial t} = \operatorname{rot} [\mathbf{v}\times\mathbf{H}],
\end{equation}
\begin{equation}
 \operatorname{div} \mathbf{H} = 0,
\end{equation}
\begin{equation}
\label{ener}
 \frac{\partial e}{\partial t} + \operatorname{div} \left( \mathbf{v} \left(e+p+\frac{\mathbf{H}^2}{8\pi} \right) - \frac{\mathbf{H}(\mathbf{v}\cdot\mathbf{H})}{4\pi}  \right) = S,
\end{equation}
\begin{equation}
\label{sost}
 p = (\gamma-1) \left( e - \frac{\rho\mathbf{v}^2}{2} - \frac{\mathbf{H}^2}{8\pi} \right),
\end{equation}
where $e=\rho\varepsilon + \frac{\rho\mathbf{v}^2}{2} + \frac{\mathbf{H}^2}{8\pi}$ is full internal energy per unit volume and $S$ the cooling rate.

The radiation cooling has been taken into account by calculating a coefficient of Planck opacity $k(\rho,T)$ in case of laboratory jet and a cooling function $\Lambda(T)$ in case of astrophysical jet calculation at each step in each cell. The coefficient $k$ is known from tables, got through a program PrOpacEOS \cite{propaceos}. The curve $\Lambda$ has been got through a CHIANTI database \cite{dere}. Values of $k$ between knots of grid $T$, $\rho$ have been calculated with  bilinear interpolation. In the approximation of an optically thin body we have the following expression for a rate of volume cooling of laboratory plasma: $S= 2k\rho\sigma T^4 $, where $\sigma$ is the Stefan Boltzmann constant. In the same approximation a rate of volume cooling of astrophysical plasma is $S= (1-\alpha)n^2\Lambda$. The applicability of this approximation is being discussed in the next paragraph.

These equations  are solved using our own numerical scheme of the Godunov type in two dimensional cylindrical coordinates using a well-proven solver HLLD \cite{kusono}. The boundary conditions have been chosen to be free, although for calculation of jet propagation a type of boundary conditions is not very important.

\subsection{Accounting for radiative cooling}

\begin{figure*}

\begin{minipage}{0.32\linewidth}
{ \includegraphics[height=1.9\linewidth]{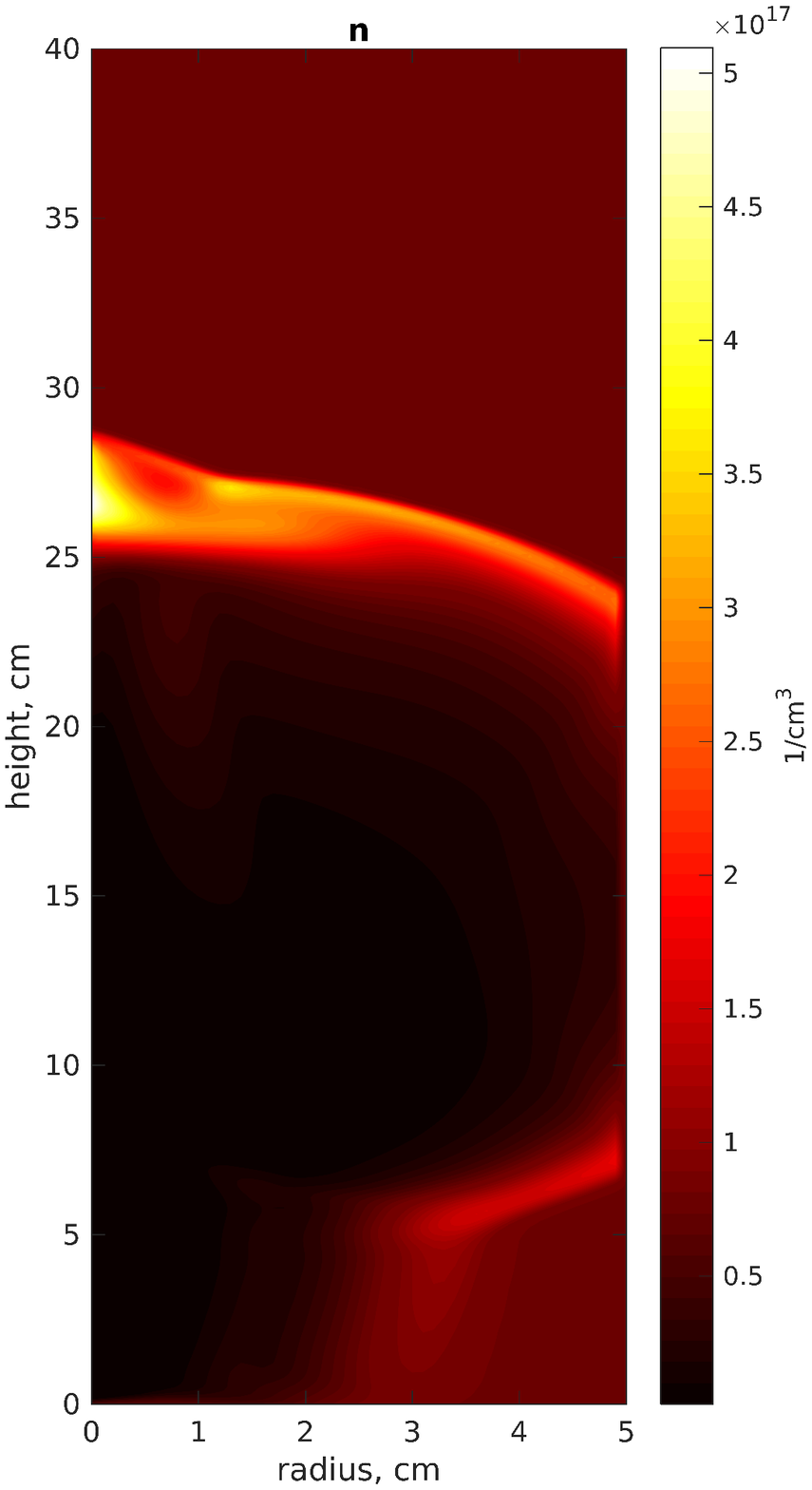} \\ a) }
\end{minipage}
\begin{minipage}{0.32\linewidth}
{ \includegraphics[height=1.9\linewidth]{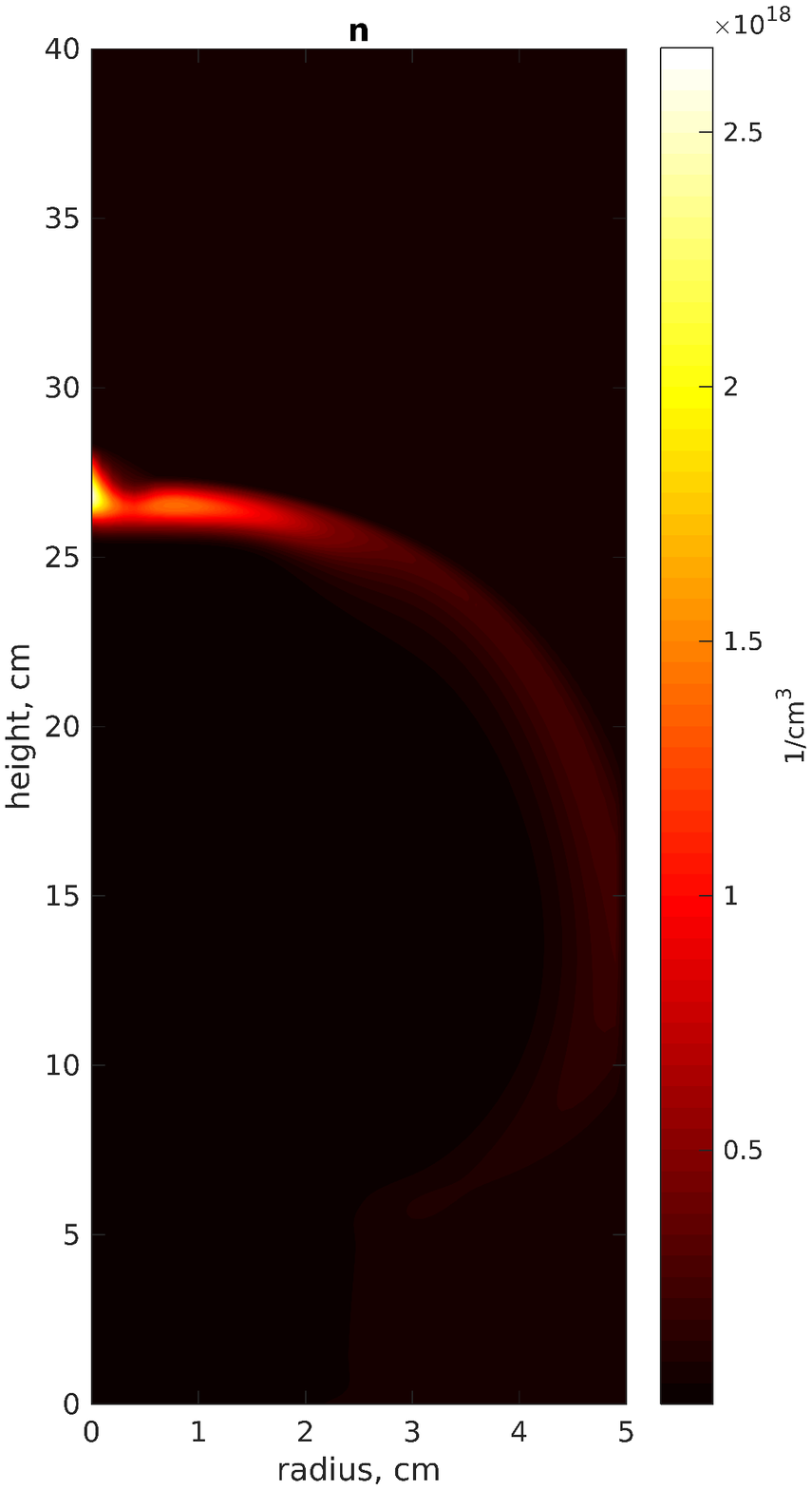} \\ b) }
\end{minipage}

\caption{Distributions of concentrations of the argon plasma in modeling the propagation of laboratory jets without - a) and with - b) taking into account the radiation cooling. }
\label{compareRad}
\end{figure*}

The optical depth of medium is $\tau=\rho k L$, where $L$ - characteristic size. For the laboratory jet it is $\tau\sim 10^{-4}$ in case of hydrogen and $\tau\sim 10^{-3}$ in case of argon, where $k$ has been calculated in the PrOpacEOS program with the most critical parameters of the plasma. To verify the assumption of a free exit of quanta from the medium under the astrophysical conditions we have calculated using Cloudy program \cite{ferland} the volume cooling rate for the observed parameters of the RW Aur's jet: $T_e=1.29\text{ eV}$, $n_H = 10^5 \text{ cm}^{-3}$, the characteristic size is a transverse section of the jet $10^{15}\text{ cm}$. With such parameters the main cooler are hydrogen lines and lines of singly ionized metals, at that optical depth for these lines is more than $1$. In order to understand how much influence of optical depth of the cooling lines on a cooling rate of the plasma the model with the same parameters but with different geometric size $10^{7}\text{ cm}$ has been calculated. With such characteristic size all lines are knowingly optical thin. So the difference in the gas cooling rate between the two models has been about $20\%$. Thus, for a thickness of $ 10^{15} \text{ cm} $ we can use the cooling curve calculated for an optical thin case, with an error not exceeding the other uncertainties of the problem (for example the metal abundances in the jet). It turns out that in all cases we can assume that the medium is optically thin and do not take into account the interaction of radiation with the medium. To account for the cooling of plasma of the astrophysical jet we have been used cooling curve $\Lambda(T)$, calculated for the solar elemental abundances\cite{grevesse} in CHIANTI database\cite{dere}, which uses the approximation of coronal equilibrium and assumes a free exit of radiation from the medium.

Because of high density of the laboratory jet populations of atomic levels are defined by collision processes, therefore the local thermodynamic equilibrium can be applied in equilibrium state. However, because of the short time of process, the distribution of argon atoms in terms of the degrees of ionization and excitation states may be subject to nonequilibrium, time-dependent effects. The study of this question is beyond the scope of this article, and we shall confine ourselves to cooling the gas calculated under the assumption of LTE. 

Let's estimate importance of the taking into account the radiating cooling. With the parameters of the laboratory plasma volume rate of cooling may achieve $S\sim 10^{10} \text{ erg/s}\text{ cm}^3$ for hydrogen and $S\sim 10^{13} \text{ erg/s}\text{ cm}^3$ for argon. It means that during its flight in the facility $t\sim 10 \mu\text{s}$ a jet loses about $10^5 \text{ erg/cm}^3$ and $10^8 \text{ erg/cm}^3$ of specific energy respectively. Wherein the specific total internal energy is $e\sim 10^6\text{ erg/cm}^3$ for hydrogen and $e\sim 10^8\text{ erg/cm}^3$ for argon. The knots of the jet considered by us have specific total internal energy about $e\sim 10^{-5} \text{ erg/s}\text{ cm}^3$ Volume rate of cooling is $S\sim 10^{-13}\text{ erg/s}\text{ cm}^3$ at the beginning moments and $S\sim 10^{-15}\text{ erg/s}\text{ cm}^3$ in the late stages, when the density of plasma knots and its temperature significantly decreased. This means that during flight time $\sim 10^9\text{ s}$ the jet loses about a tenth of the total internal energy through radiation. Comparison of the movement dynamic of the emitting and non-emitting jets has not revealed significant differences, however, in the future, for the sake of completeness, we will take into account the radiative cooling of astrophysical jet.

\begin{figure*}

\begin{minipage}{0.32\linewidth}
\center{ \includegraphics[height=1.9\linewidth]{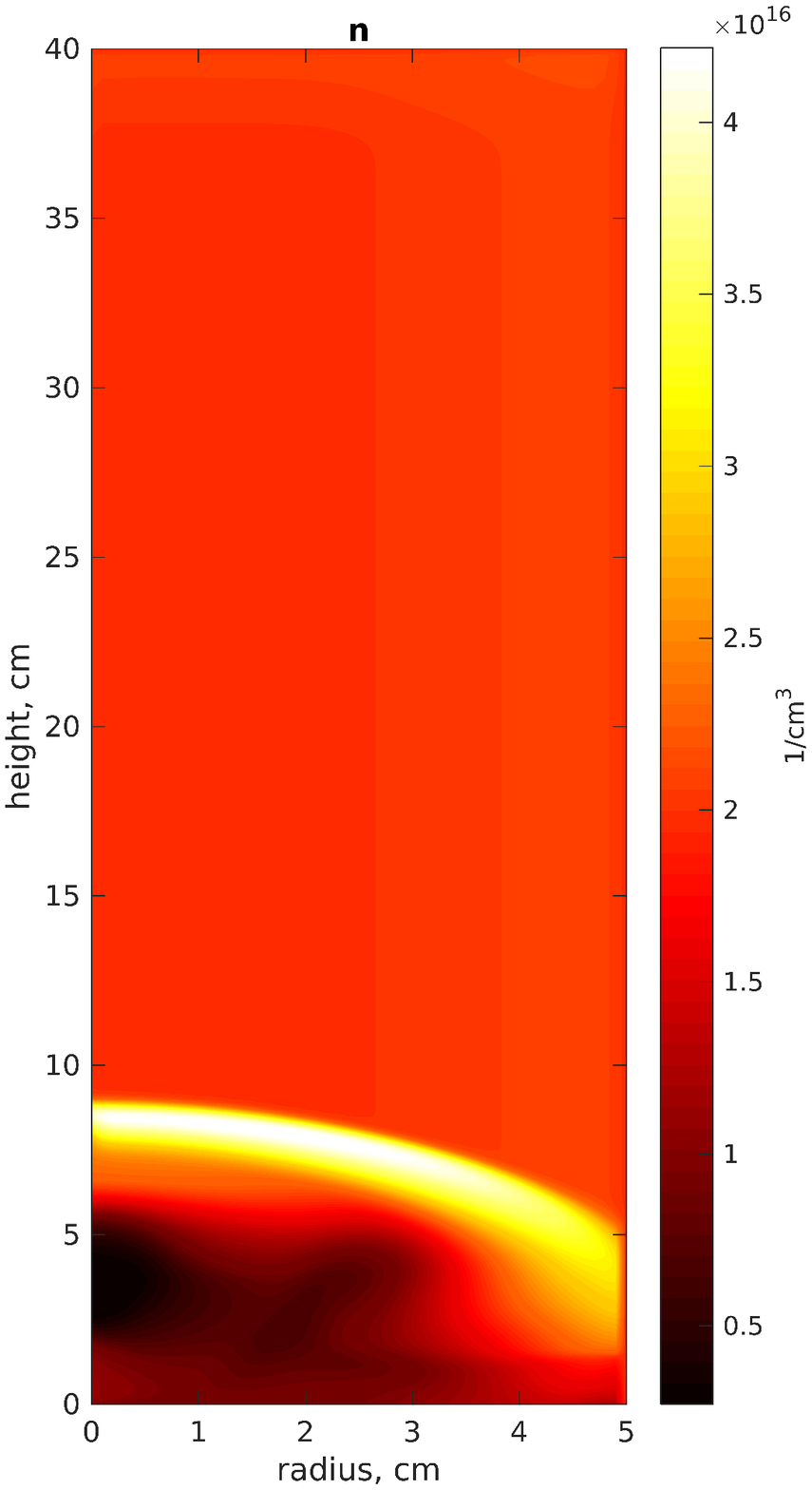} \\ a) }
\end{minipage}
\begin{minipage}{0.32\linewidth}
\center{ \includegraphics[height=1.9\linewidth]{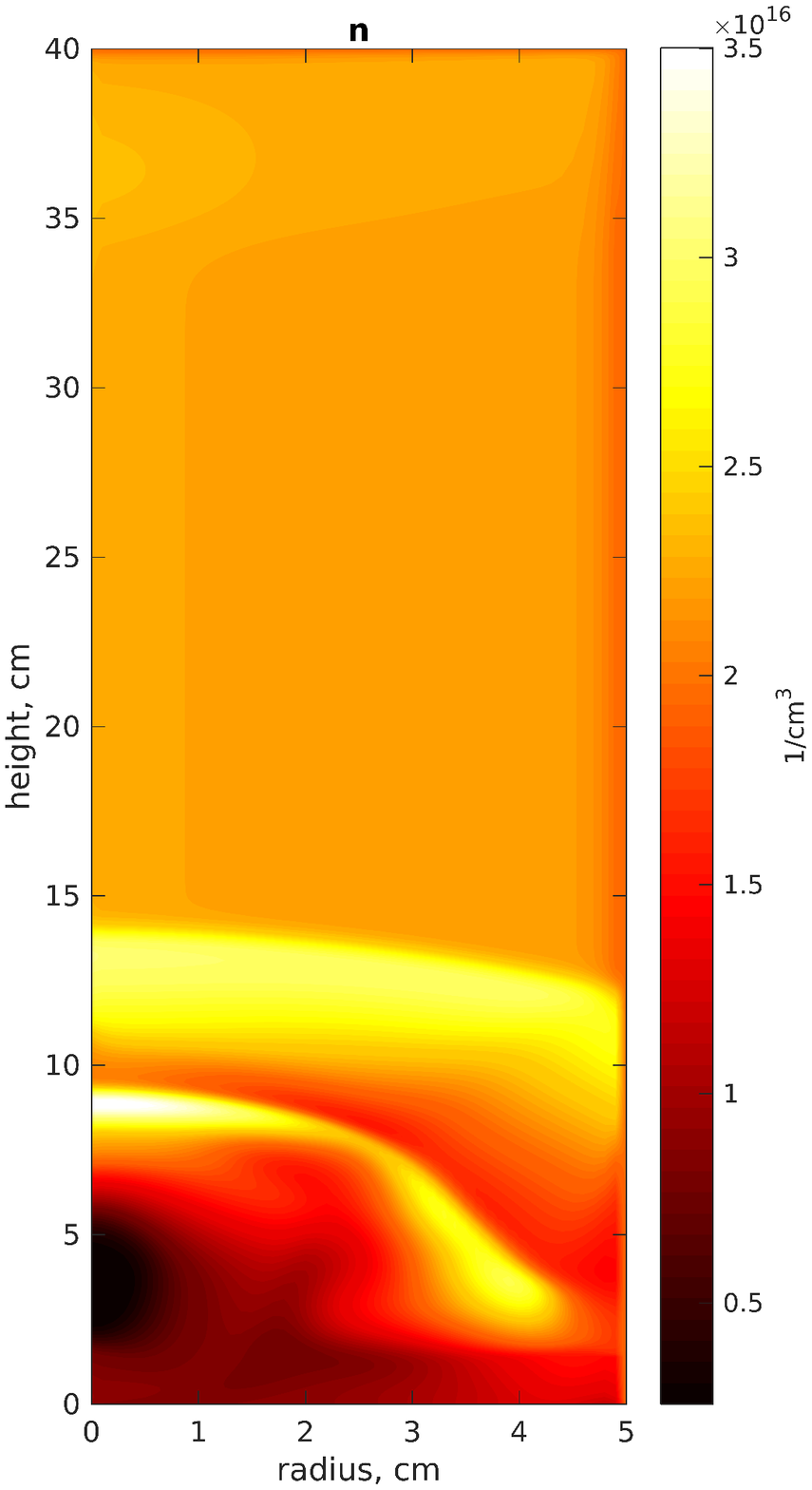} \\ b) }
\end{minipage}
\begin{minipage}{0.32\linewidth}
\center{ \includegraphics[height=1.9\linewidth]{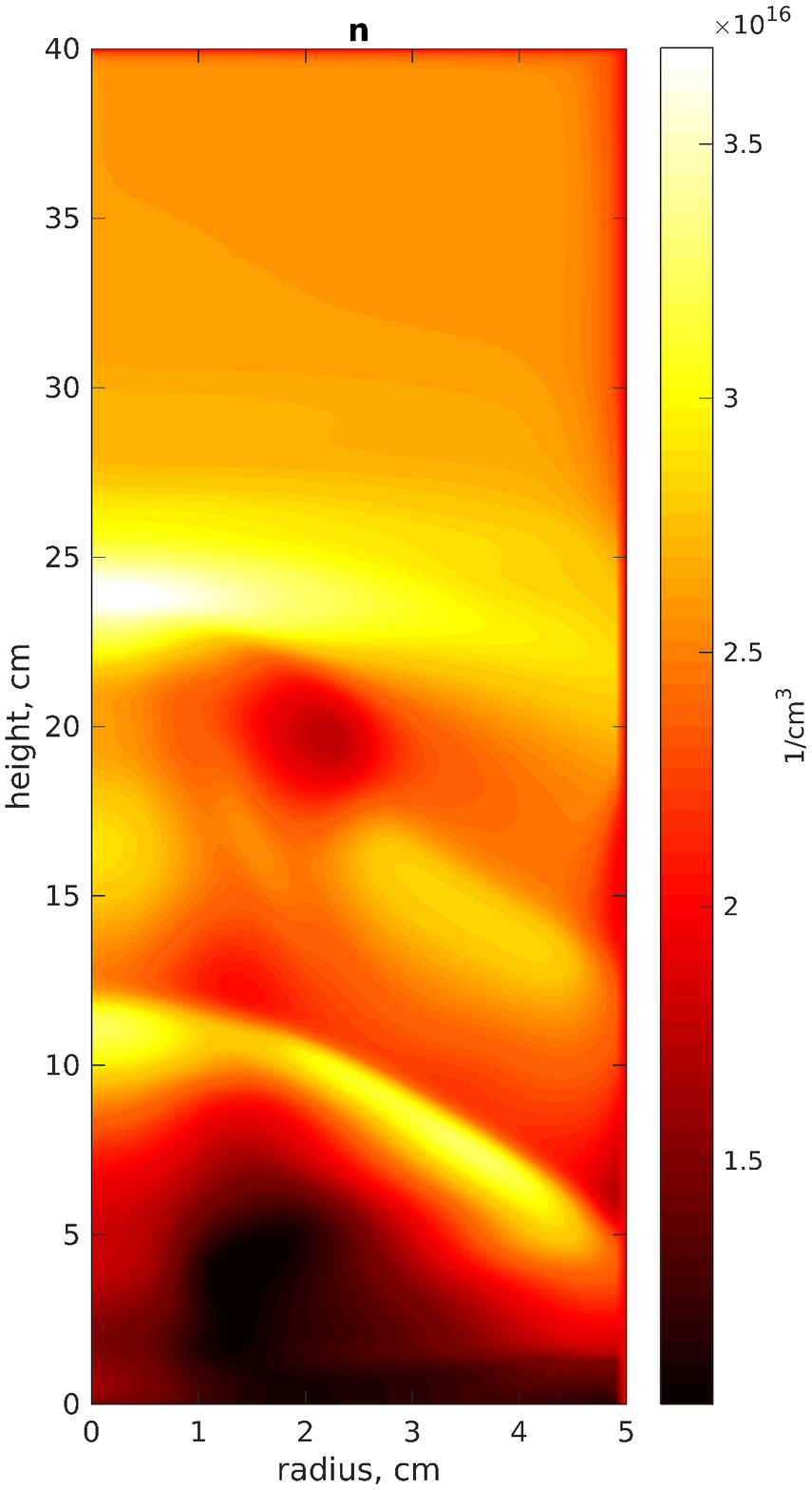} \\ c) }
\end{minipage}

\caption{Distribution of hydrogen plasma concentrations in modeling the propagation of laboratory jets at time points: a) $2\mu\text{s}$, b) $5\mu\text{s}$ and c) $18\mu\text{s}$.}
\label{H_3mcs}
\end{figure*}

It turns out that during the flight the laboratory jet of argon loses a significant part of its internal energy due to radiation. It leads to decreasing of pressure within the jet, which contributes to its collimation. Thus value $S$ in (\ref{ener}) is not small and makes a significant contribution into energy balance equation and has a significant influence on the dynamics of the propagation of argon jets. But for greater accuracy we have taken into account the radiative cooling of all types of jets.

Let's compare results of modeling, carried out with typical parameters of jets, formed during the experiments with argon, taking into account the radiative cooling and without it, with other equal parameters. As initial conditions the following parameters have been taken: $n_\text{jet} = 4\cdot 10^{17} \text{ cm}^{-3}$, $n_\text{ambient} = 8\cdot 10^{16} \text{ cm}^{-3}$, $T_\text{jet}= 1.5 \text{ eV}$, $T_\text{ambient}= 1 \text{ eV}$, $V_\text{jet} = 5\cdot 10^6 \text{ cm/s}$, $B_0 = 4.5 \text{ kG}$. It can be seen from fig. \ref{compareRad}, that both matter in the central knot and mushroom-like shock wave are more pressed to the axis in case of radiative jet (b), than non-radiative jet (a). Wherein the emitting jet has five times greater density. This clearly demonstrates the importance of radiation cooling for the collimation of the jet.

\section{Results and discussion}

\subsection{An example of a laboratory jet}

In previous work \cite{Kalash}, we have  studied the propagation of a single plasma knot. Let summarize the main results. We have concluded, that presence of toroidal magnetic field prevents spillovering of matter from a jet to a shock wave. Apparently the collimation of laboratory jets occurs solely due to the magnetic field. 
Also, the formation of a region with a low density and an elevated plasma temperature was observed. This region remains after passing of a jet. A comparison between jets with different parameters shows the importance of the ratio of the densities of the jet and its ambient. The larger the ratio, the less resistance the jet experiences and the less matter appears in the mushroom-like shock wave.

\begin{table}
\caption{\label{tabl:Init} The parameters of the initial conditions.}
\begin{center}
\begin{tabular}{lcc}
\hline\hline
 \ {\footnotesize Parameter }		&\ {\footnotesize Laboratory }	&\ {\footnotesize Astrophysical } \\
\hline
\
\ {\footnotesize Jet density, $n_\text{jet}$, $\text{cm}^{-3}$ } &\ {\footnotesize $8\cdot 10^{16}$} &\ {\footnotesize$ 65900$ }\\
\ {\footnotesize Ambient density, $n_\text{ambient}$, $\text{cm}^{-3}$ } &\ {\footnotesize $2.5\cdot 10^{16}$} &\ {\footnotesize $2200$ \& $950$  } \\
\ {\footnotesize Jet temperature, $T_\text{jet}$, $\text{eV}$ } &\ {\footnotesize$2$} &\ {\footnotesize$1.06$} \\
\ {\footnotesize Ambient temperature, $T_\text{ambient}$, $\text{eV}$ } &\ {\footnotesize$1.1$} &\ {\footnotesize$0.26$} \\
\ {\footnotesize Maximum magnetic field, $B$, $\text{kG}$ } &\ {\footnotesize$4.5$} &\ {\footnotesize$0$} \\
\ {\footnotesize Velocity, $V_\text{jet}$, $\text{cm/s}$ } &\ {\footnotesize$5\cdot 10^6$} &\ {\footnotesize$1.5\cdot 10^7$} \\
\ {\footnotesize Launch frequency } &\ {\footnotesize$3\mu\text{s}$}  &\ {\footnotesize$26$  $\text{yrs}$} \\

\hline\hline
\end{tabular}
\end{center}
\end{table}

In this new series of tests, we have analyzed the propagation of a succession of plasma knots emitted at periodic interval of time of $3\mu\text{s}$. The initial conditions for the  parameters of the plasma in PF-3 are 
presented in tabl. \ref{tabl:Init}. With this choice of initial conditions, the parameters observed in the laboratory are achieved by modeling after several steps in time. The geometry of such a formulation and the magnetic field configuration described above and the gas used are hydrogen and argon.

The figure \ref{H_3mcs} shows the results of this simulation for hydrogen. The propagation of the plasma knots  are represented at different times $3\mu\text{s}$ , $5\mu\text{s}$ and $18\mu\text{s}$ as the function of the density.  In fig. \ref{H_3mcs}a the first knot moves without collimation  and the classical mushroom-like shock wave is visible. The matter is distributed almost uniformly. In fig 2 b, the second knot emitted at $3\mu\text{s}$, has almost the same structure for the shock.
Finally the last figure \ref{H_3mcs}c at $18\mu\text{s}$ shows again no collimation and a behavior identical the the previous knots. We conclude that in this simulation, hydrogen knots propagate without collimation.

\begin{figure*}

\begin{minipage}{0.32\linewidth}
\center{ \includegraphics[height=1.9\linewidth]{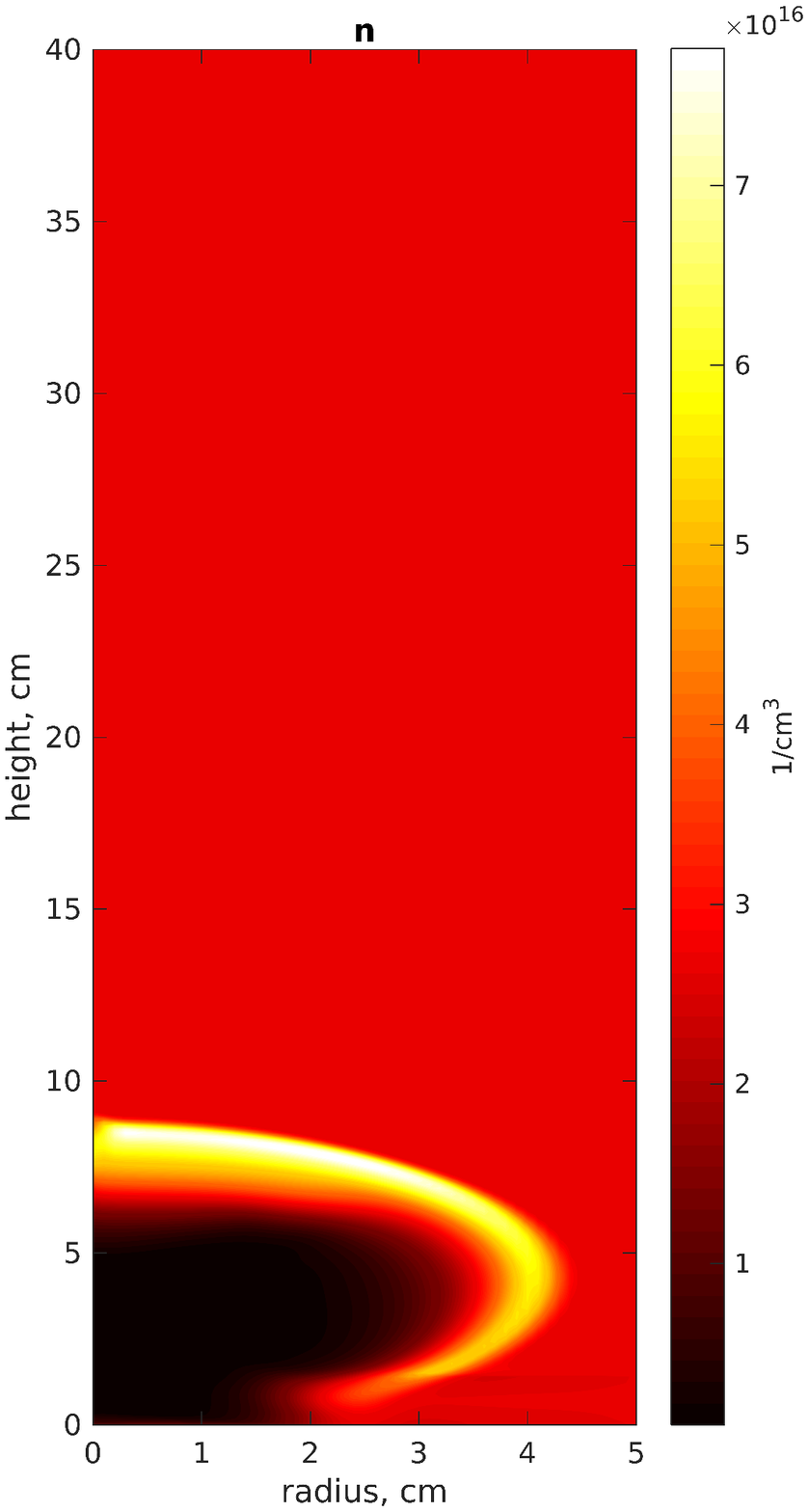} \\ a) }
\end{minipage}
\begin{minipage}{0.32\linewidth}
\center{ \includegraphics[height=1.9\linewidth]{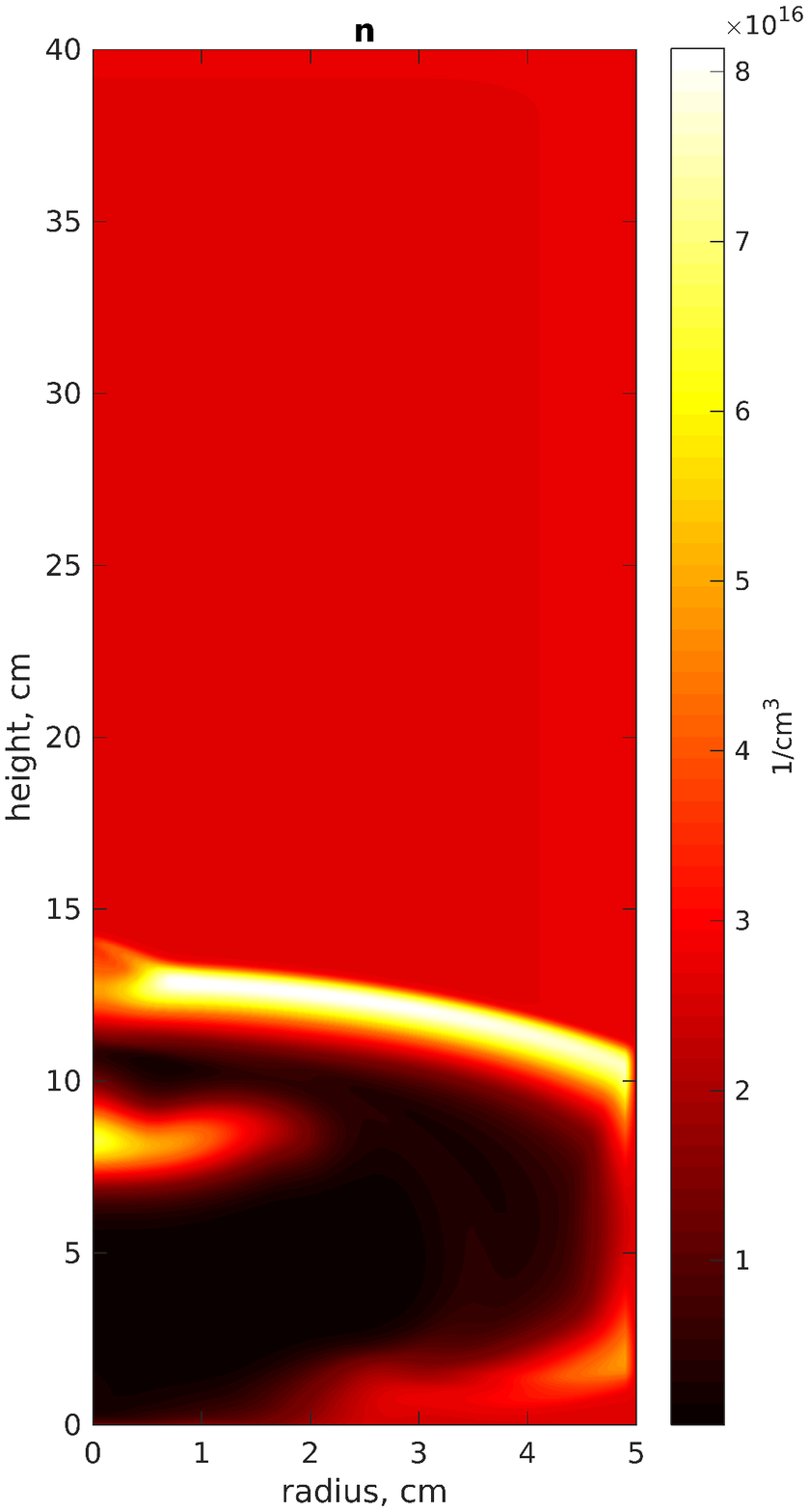} \\ b) }
\end{minipage}
\begin{minipage}{0.32\linewidth}
\center{ \includegraphics[height=1.9\linewidth]{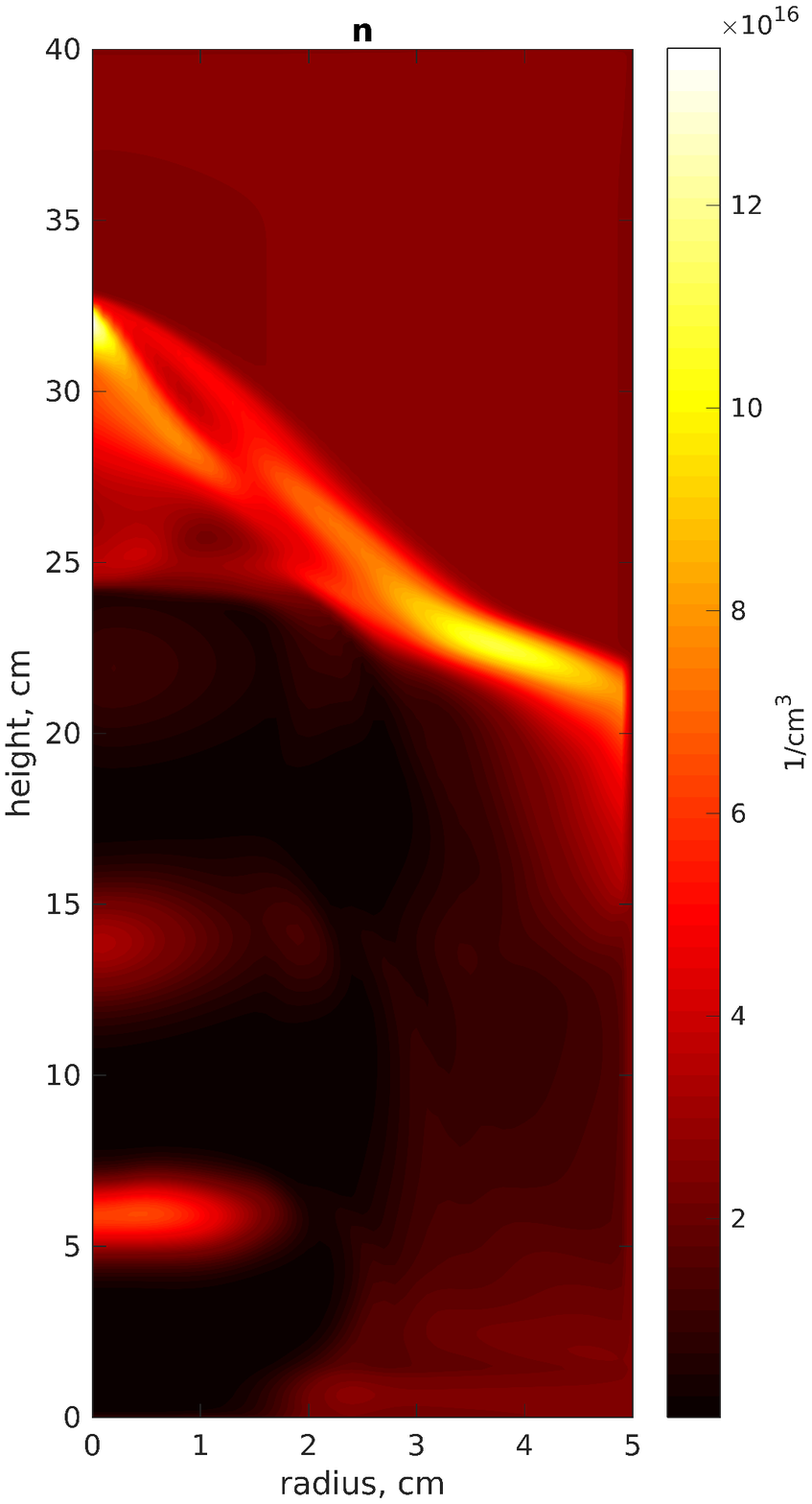} \\ c) }
\end{minipage}

\caption{Distribution of argon plasma concentrations in modeling the propagation of laboratory jets at time points: a) $2\mu\text{s}$, b) $6\mu\text{s}$ and c) $20\mu\text{s}$.}
\label{Ar_3mcs}
\end{figure*}

The second simulation regards the case of argon plasma, which is being emitted also each $3\mu\text{s}$. As it follows from fig. \ref{Ar_3mcs}a, the shock wave of the first ejection has smaller radius than hydrogen one, but it propagates without collimation too, because all plasma of initial knot is also distributed over the shock wave homogeneously. However the next knot (fig. \ref{Ar_3mcs}b) at the same distance has very weak shock wave and the overwhelming majority of the plasma remains within the initial radius $1.5 \text{ cm}$. Since the knot following the first turn out to be in a region with low density, they experience less resistance from the ambient, whereby they are slowed down less. Therefore they overtake the shock wave, formed by the first ejection and interact with it, forming a puzzled structure of interacting shock wave (fig. \ref{Ar_3mcs}c). But our aim is investigation of influence of channel with low density plasma, formed by the first ejection on collimation of subsequent ejections. And, as can be seen from fig. \ref{Ar_3mcs}c, the radial size of these ejections increases very little, apparently due mainly to thermal expansion with the speed of sound (see below).

\subsection{Modeling an astrophysical jet}

\begin{figure*}

\begin{minipage}{0.32\linewidth}
\center{ \includegraphics[height=1.9\linewidth]{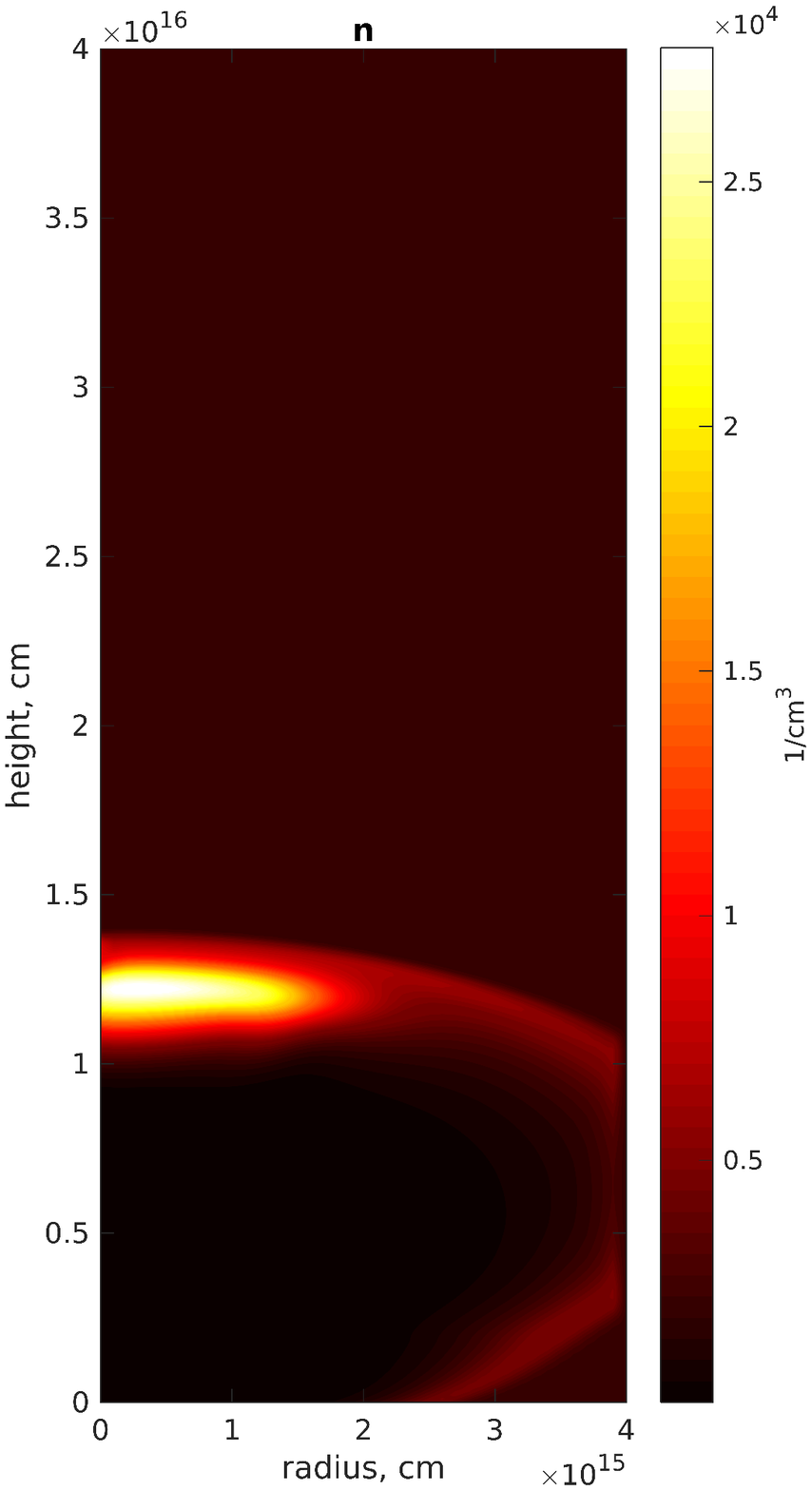} \\ a) }
\end{minipage}
\begin{minipage}{0.32\linewidth}
\center{ \includegraphics[height=1.9\linewidth]{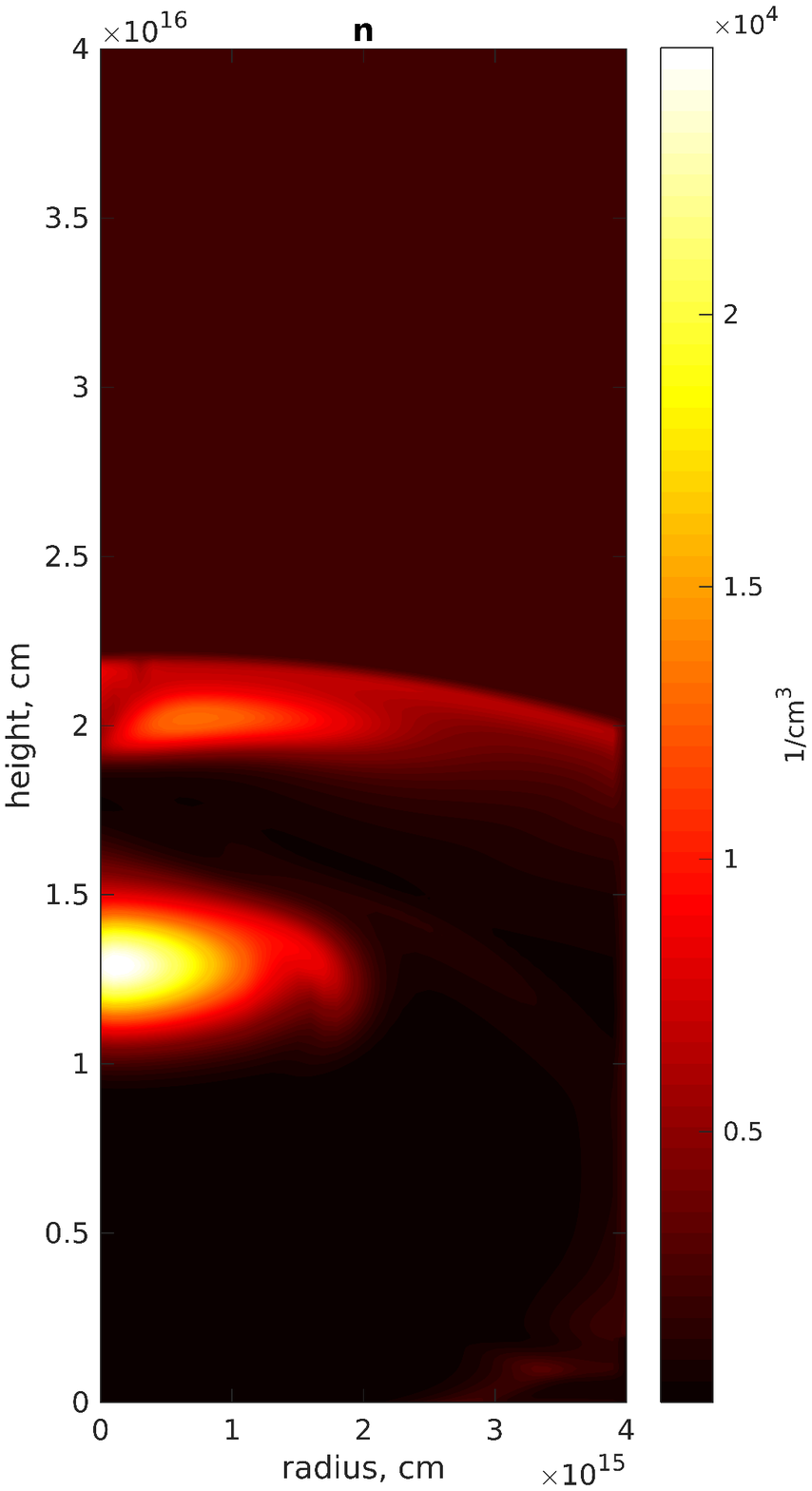} \\ b) }
\end{minipage}
\begin{minipage}{0.32\linewidth}
\center{ \includegraphics[height=1.9\linewidth]{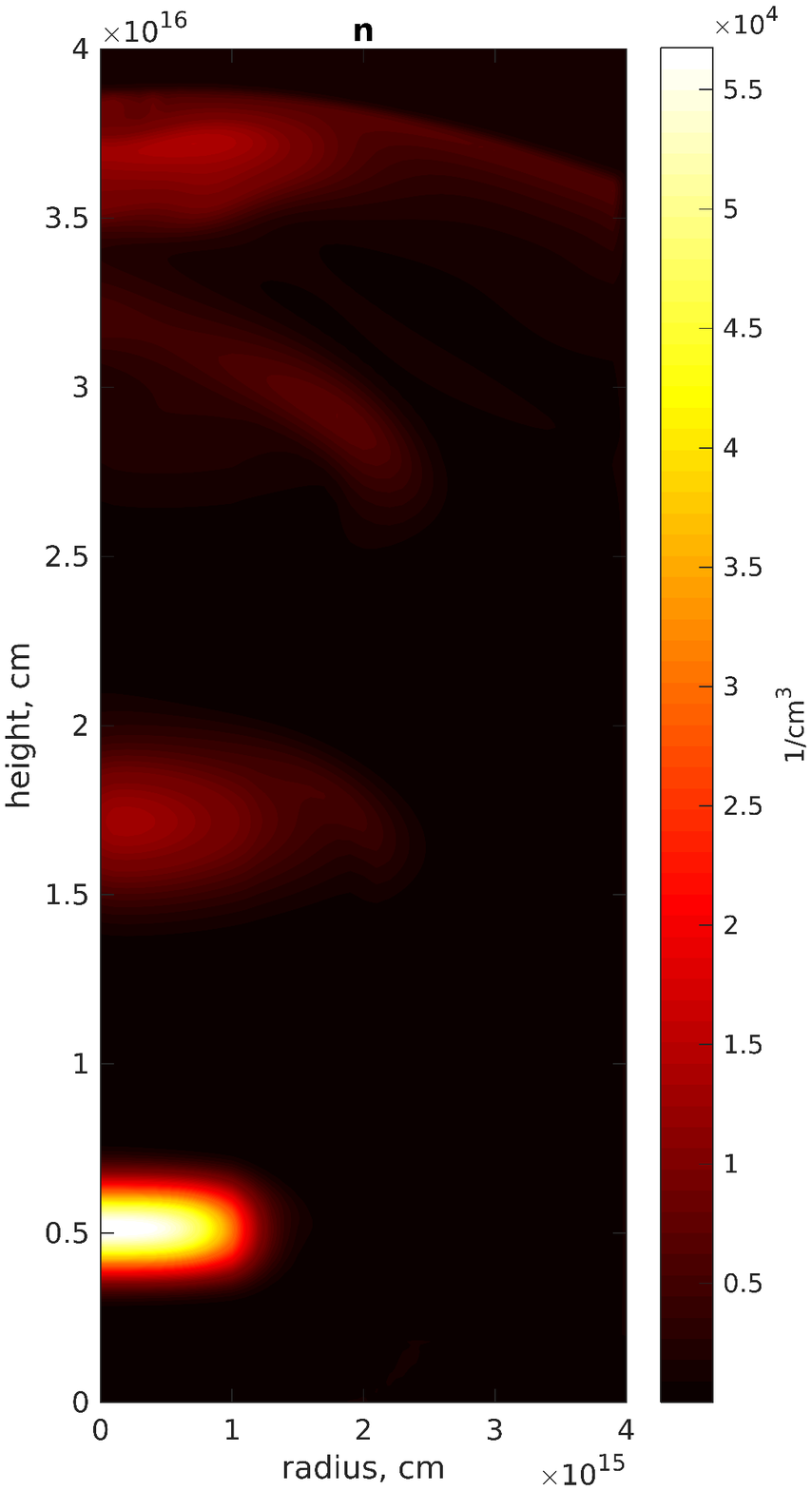} \\ c) }
\end{minipage}

\caption{Distribution of plasma concentrations in modeling the propagation of astrophysical jets with density contrast $n_\text{jet}/n_\text{ambient}=30$ at time points: a) $25.7\text{ yrs}$, b) $51.1\text{ yrs}$ and c) $110.7\text{ yrs}$.}
\label{astro30}
\end{figure*}

The astrophysical numerical simulation of jet in YSOs, we have chosen the red (i.e. propagating from an observer) jet of T Tauri star RW Aur. The geometrical characteristics of this jet are known very accurately, as well as some physical ones, such as electron density, ionization degree, temperature, axial velocity and flow velocity\cite{Astro}, therefore we have chosen the parameters of initial knot
, presented at tabl. \ref{tabl:Init}. Radius of the knot is $r=10^{15}\text{ cm}$ and length is $l=2\cdot 10^{15} \text{ cm}$. We know nothing about properties of the ambient plasma, therefore we have considered two cases, when the density contrast between the knot and ambient is $30$ and $70$, but ambient temperatures are the same. Since we do not have any information about either a value or a structure of the magnetic field and also our configuration of magnetic field has only its azimuthal component, supporting to collimation, we have done the calculations without magnetic field. It demonstrates better the described effect of collimation of knot, following by the first one. Identical plasma knots are supposed to appear in the place of the first one each $26$ years, which is connected with observations for RW Aur star.

\begin{figure*}

\begin{minipage}{0.32\linewidth}
\center{ \includegraphics[height=1.9\linewidth]{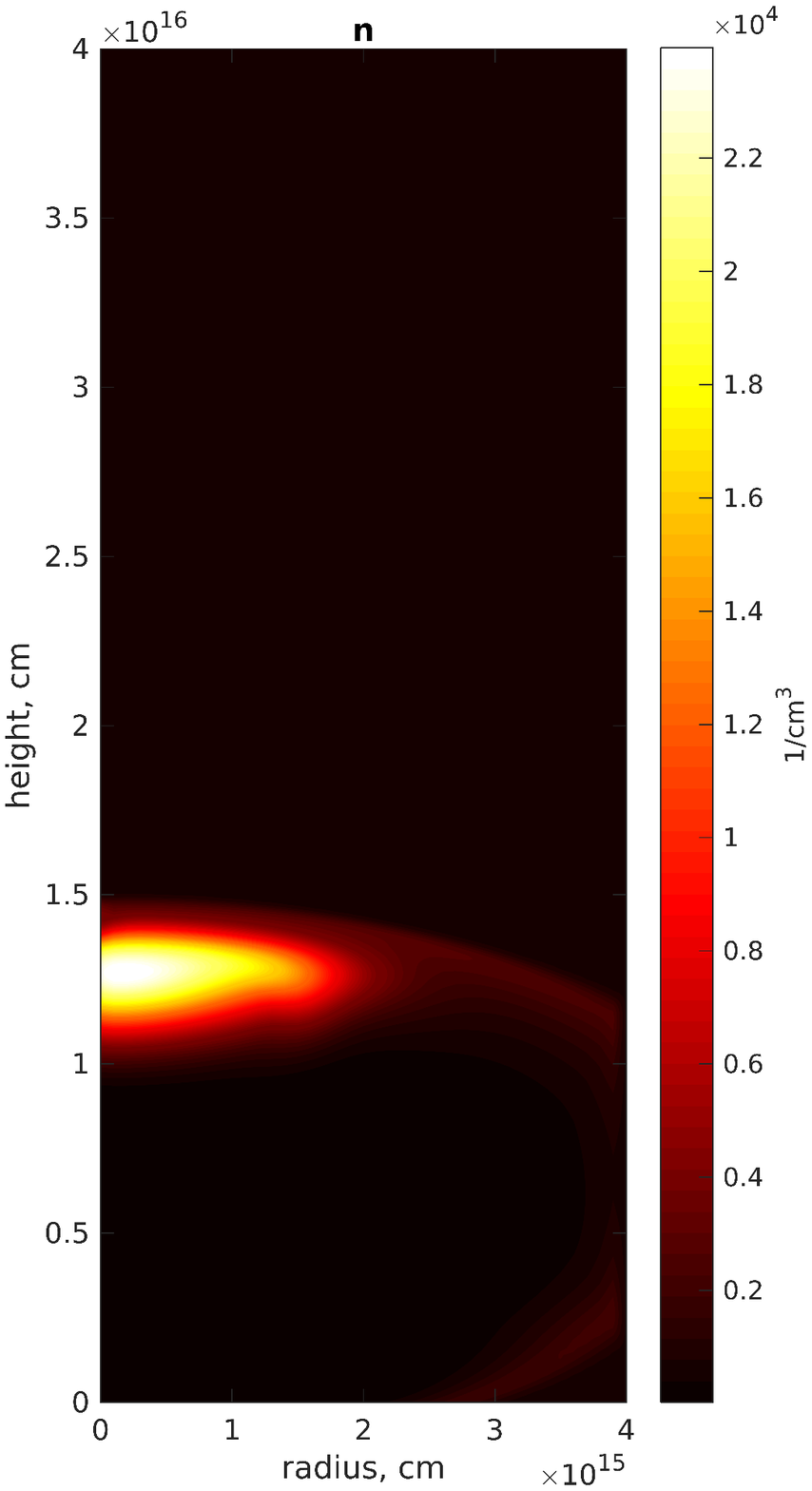} \\ a) }
\end{minipage}
\begin{minipage}{0.32\linewidth}
\center{ \includegraphics[height=1.9\linewidth]{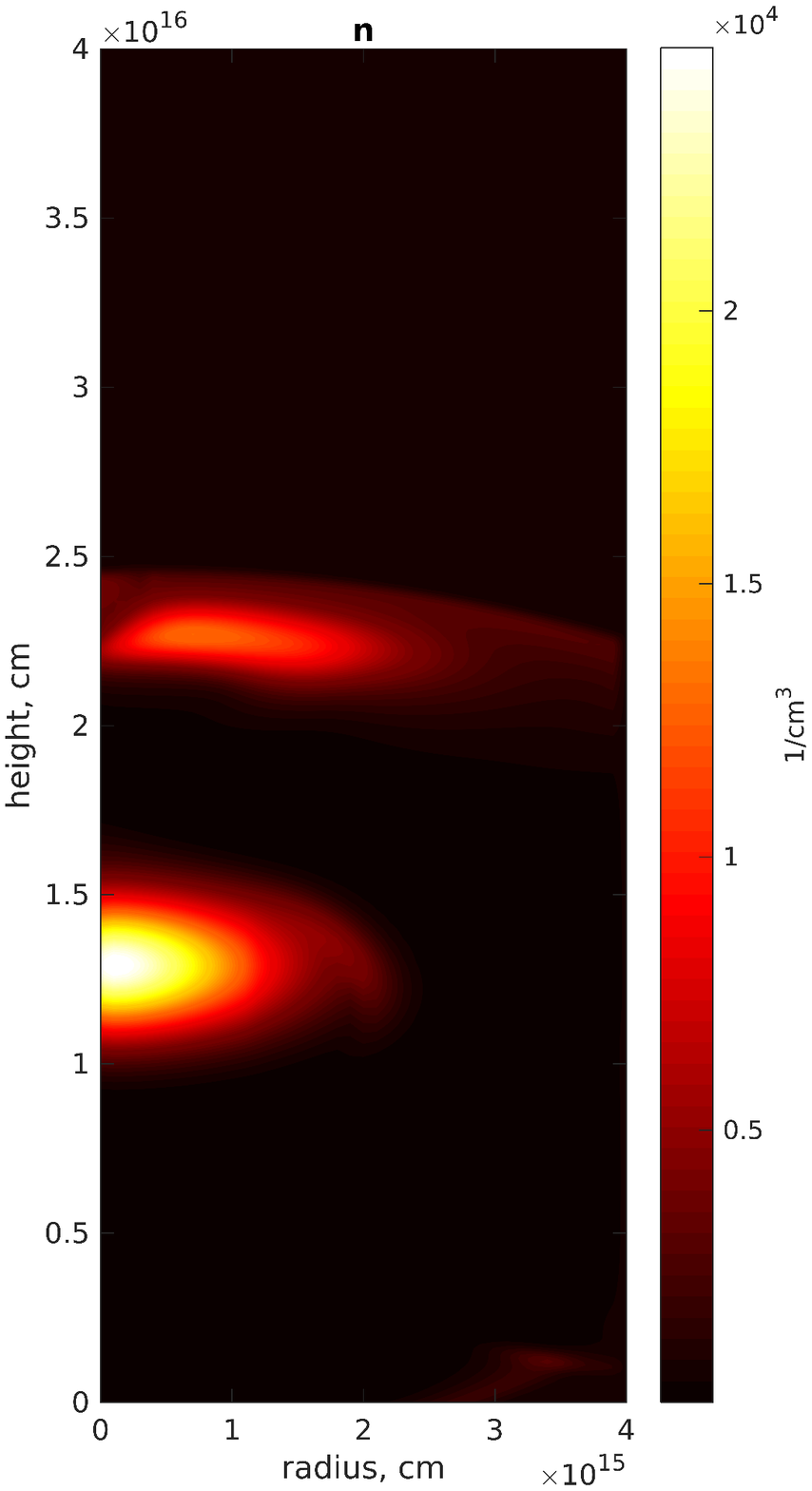} \\ b) }
\end{minipage}
\begin{minipage}{0.32\linewidth}
\center{ \includegraphics[height=1.9\linewidth]{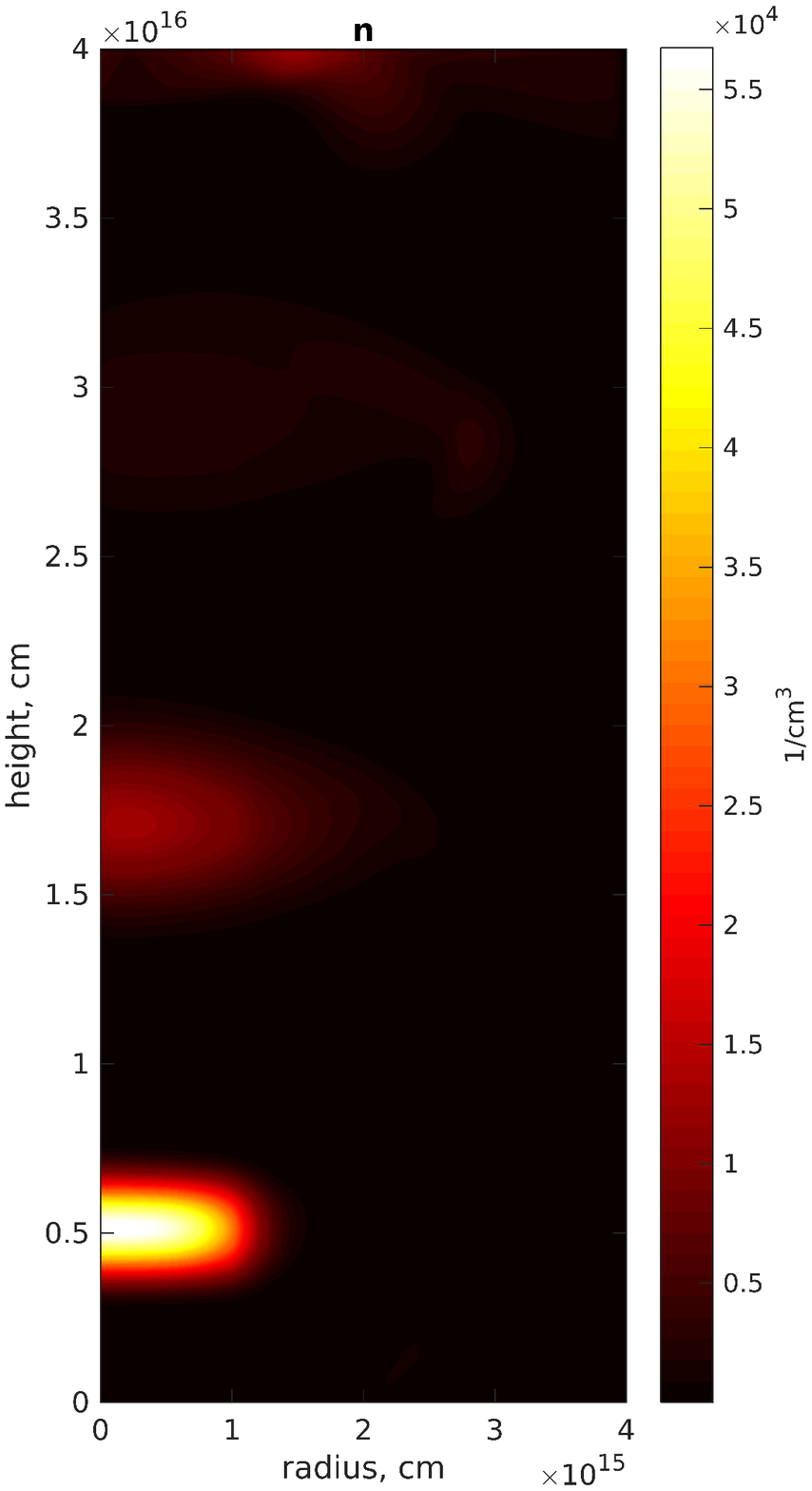} \\ c) }
\end{minipage}

\caption{Distribution of plasma concentrations in modeling the propagation of astrophysical jets with density contrast $n_\text{jet}/n_\text{ambient}=70$ at time points: a) $25.7\text{ yrs}$, b) $51.1\text{ yrs}$ and c) $110.7\text{ yrs}$.}
\label{astro70}
\end{figure*}

The results of our computations  are presented in figs. \ref{astro30} and \ref{astro70}. Analysing them it can be concluded that a jet propagates more collimated in rarefied medium. Indeed, in fig. \ref{astro70}a the plasma is more pinned to the $z$ axis, than in \ref{astro30}a, in which the more significant part of the matter flowed into the mushroom-like shock wave. This result has been got by us in the work \cite{Kalash}.
Then let's compare the shape of the first knot, passed $1.3\cdot 10^{16}\text{ cm}$ and the next one, located at the same distance after $25.4\text{ yrs}$ i.e. figs \ref{astro30}a and \ref{astro30}b also figs. \ref{astro70}a and \ref{astro70}b. We can see that the second plasma knot was destroyed much less, than the first one at the same place. Herewith the second knot has weaker shock wave than the first. It should also be noted that the second ejection spreads in a more rarefied ambient than the first one, so it experiences a lower resistance. Looking at figs. \ref{astro30}c and \ref{astro70}c it may be seen that no one knot was destroyed as the first one, which was completely transformed into the mushroom-like shock wave at the end of the calculations. 
The absence of collimation is probably due to the  matter spillover from initial knot to mushroom-like shock wave. For subsequent knots, this effect is much weaker than for the first one because the area with low density remains after passing of the first ejection. Therefore the collimated motion of whole jet is caused by presence of channel with low density and low pressure, formed by the first ejection.

\subsection{Collimation of jets}
  
According to our simulations, we found that the collimation of the jets (in laboratory conditions or in astrophysical environment) is due to the low density region   created by the propagation of the knots.  

Let's estimate the time, necessary for filling the area remained after the first ejection passing. Consciously overestimating the estimate, suppose that this cavity does not contain any substance, i.e. is a vacuum. The surrounding plasma must move toward this cavity with the maximum possible in this case speed - the speed of sound: $v=c_s\sqrt{2/(\gamma-1)} \simeq c_s$. An area with size $R_0$ must be filled in time $\tau\simeq R_0/c_s = M R_0/V$. For considering astrophysical jets this time is about $70$ years. Since the new ejections appear approximately each $25$ years then we can say that the vacuum cavity does not fill up and each new plasma knot moves within more rarefied medium, than within a surrounding cloud. For laboratory jets the time of filling is about $10\mu\text{s}$. I.e. for reproducing of this effect under a laboratory condition it is necessary that new plasma ejections appear more often than once in $10\mu\text{s}$.

Suppose now that due to this effect, a very weak mushroom-like shock wave is formed in subsequent clusters and the most part of knots  does not spillover into it. That is, we shall neglect the leakage of matter into the shock wave. Then all the movement of a jet is superposition of translational motion along the axis of rotation of a compact object and an expansion with a velocity of the order of the sound speed in the jet. Here we again exceed our estimate, assuming for the sake of simplicity that the jet expands into vacuum. Then an opening angle of the jet may be expressed as $\tan\alpha = { R(t)/L(t) }$, where $R(t)\simeq R_0 + c_s t$ is evolution of knot radius with time, $c_s$ is internal speed of sound and $L(t)=Vt$ is path, passed by a knot. Expressing it through the internal Mach number we can get:
\begin{equation}
 \alpha = 2\arctan \left( M^{-1}_* + \frac{R_0}{L} \right).
\end{equation}

For laboratory jets this number is $\alpha_H \approx 80^\circ$ for hydrogen and $\alpha_{Ar} \approx 38^\circ$ for argon. These enough big numbers confirm the got in \cite{Kalash} result, that without magnetic field a collimated motion of a jet does not occur in PF-3 facility. Also it can be noticed that $\alpha_H > \alpha_{Ar}$, which also agrees with the observations \cite{Krauz1,Krauz3}. As can be seen from figs. \ref{astro30} - \ref{astro70}, the following for the first outbursts of the RW Aur star just as well fits into this dependence, despite the fact that the initial size of the knot was chosen to be smaller than the emissions near the star, which are unknown to us from observations.

\section{Conclusions}

Having carried out numerical simulation of a hypothetical situation in which the plasma emissions in the PF-3 unit can occur one by one we have seen that a cavity with a low concentration of plasma formed after the first knot  prevents the formation of a strong shock wave: the successive knots propagate without mushroom effect and the jet remains collimated. In this way the most part of matter remains the within initial boundaries of an emission. The knot itself expands, apparently, because of the thermal expansion into the environment. Carrying out a similar simulation with parameters of the observed jet of the RW Aur star, we have found the same effect.

A question about a possibility of an experimental verification of the results of  the numerical simulation on the PF-3 facility remains open. Up to now, the main attention in the conducted experiments was paid to the study of the head clot. However, the results of experiments show that in plasma-focus devices, several successive compressions of the current sheath on the axis occur and, consequently, the generation of consecutive clots separated by time from hundreds nanoseconds to several microseconds \cite{Krauz1,Krauz6}. The main problem is that this process is difficult to control and the parameters of the bunches generated in different compressions may initially differ from each other. Nevertheless, it may be an important area for our further experimental studies.

Further, making an estimate of the scattering angle of plasma knots, we have assumed that their motion is a superposition of translational motion along the initial direction of propagation and thermal expansion into vacuum. Although this estimate is certainly overestimated, since neither the environment nor the magnetic field was taken into account, it is still qualitatively consistent with the results of modeling and astronomical observations.

Of course, the jets propagate collimated not only due to the vacuum region formed by the first ejection. An important role is played also by magnetic fields, and radiation and pressure of the environment and other effects, possibly not known to us.

\section*{acknowledgments}
The authors are grateful to E.P. Velikhov for the formulation of the problem, A.V. Shilkov, V.S. Beskin and S.A. Lamzin for useful discussions.

The work related to the laboratory installation PF-3 was carried out with the support of the RFBR grant 17-02-01184 A. The work related to the modeling of the astrophysical jet was carried out with the support of the RSF grant 16-11-10339. I. Kalashnikov thanks the joint French-Russian Ph.D. program in Fundamental Physics for financial support.

\end{document}